\documentclass[%
 reprint,
 amsmath,amssymb,
 aps,
]{revtex4-1}

\usepackage{graphicx}
\usepackage{dcolumn}
\usepackage{bm}

\usepackage{caption}
\usepackage{subcaption}
\usepackage{mhchem}

\begin{document}

\preprint{APS/123-QED}

\title{Normal State Correlates of Plasmon-Polaron Superconductivity in Strontium Titanate}

\author{Alexander Edelman}
\email{aoe@uchicago.edu}
\author{Peter Littlewood}
\affiliation{James Franck Institute and Department of Physics, University of Chicago, IL 60637, USA}
\date{\today}

\begin{abstract}
We analyze a minimal model of \ce{SrTiO3}, such as has been historically used to analyze the superconducting state. Treating the electron-phonon and electron-electron interactions on an equal footing, we calculate the normal state properties, focusing on the spectral function. We find a density-driven crossover caused mostly by the adiabaticity of the bosonic modes, but the spectral features of our theory are dominated by coupled plasmon-phonon oscillations rather than the bare phonon features seen in experiment.
\end{abstract}
\maketitle

Strontium titanate, \ce{SrTiO3}, is a cubic perovskite wide band gap semiconductor which at low temperatures approaches a ferroelectric phase transition but instead saturates at a dielectric constant $\epsilon_0 \sim 2\times10^4$, reflecting the presence of an extremely soft transverse optic (TO) phonon\cite{Muller:1979je,Coak_Haines_Liu_Rowley_Lonzarich_Saxena_2018}. Against this background of quantum criticality, when lightly doped the material hosts a plethora of interesting phenomena, including Fermi liquid-like resistivity above the Fermi temperature\cite{Lin_Rischau_Buchauer_Jaoui_Fauque_Behnia_2017}, phonon hydrodynamics\cite{Martelli_Jimenez_Continentino_Baggio-Saitovitch_Behnia_2018}, and a superconducting dome spanning electron densities from $10^{17}$ to $10^{21}\text{cm}^{-3}$\cite{Schooley_Hosler_Cohen_1964, Lin_Zhu_Fauque_Behnia_2013, Bretz-Sullivan_Edelman_Jiang_Suslov_Graf_Zhang_Wang_Chang_Pearson_Martinson_2019}, with a transition temperature peaking around $300\text{mK}$, on the order of $1\%$ of the Fermi energy ($E_F$), comparable to the high-$T_C$ materials. The superconducting state nevertheless appears to be $s$-wave and weakly coupled, with a BCS-like $\Delta/T_C$ ratio\cite{Thiemann_Beutel_Dressel_Lee-Hone_Broun_Fillis-Tsirakis_Boschker_Mannhart_Scheffler_2018,Yoon_Swartz_Harvey_Inoue_Hikita_Yu_Chung_Raghu_Hwang_2021}. 

It has long been known that conventional pairing by acoustic phonons cannot explain the observed behavior, and superconductivity in \ce{SrTiO3} has thus spurred great theoretical creativity, both historically and very recently\cite{Koonce_Cohen_Schooley_Hosler_Pfeiffer_1967, Gastiasoro_Trevisan_Fernandes_2020, Volkov_Chandra_Coleman_2021, Kanasugi_Kuzmanovski_Balatsky_Yanase_2020, Wolfle_Balatsky_2018, Dunnett_Narayan_Spaldin_Balatsky_2017,Arce-Gamboa_Guzman-Verri_2018, Kedem_Zhu_Balatsky_2016, Marel_Barantani_Rischau_2019}. Here we will focus on a particular class of theories, pioneered by Takada\cite{Takada_1980}, which involve the electron gas coupling to a longitudinal optic (LO) phonon mode \cite{Klimin_Tempere_Devreese_Marel_2016,Rowley_Enderlein_Oliveira_Tompsett_Saitovitch_Saxena_Lonzarich_2018, Ruhman_Lee_2016,Gastiasoro_Chubukov_Fernandes_2019, Gorkov_2016}. These theories face the difficulty that the coupling to this mode is of the same long-range Coulomb character as the repulsion between the electrons, and the two effects must therefore be treated on an equal footing. An additional problem as that the LO frequency $\Omega \gg E_F$. As pointed out by \cite{Swartz_Inoue_Merz_Hikita_Raghu_Devereaux_Johnston_Hwang_2018}, a naive calculation of the BCS coupling constant due to the LO phonon far exceeds the value measured in the superconducting state. Although the elecron-phonon coupling strength in the sense of the Fr{\"o}lich coupling constant $\alpha\sim 2$ \cite{Devreese_Klimin_Mechelen_Marel_2010} is intermediate at best, both photoemission \cite{Wang_Walker_Tamai_Wang_Ristic_Bruno_Torre_Ricco_Plumb_Shi_2016} and  tunneling\cite{Yoon_Swartz_Harvey_Inoue_Hikita_Yu_Chung_Raghu_Hwang_2021,Swartz_Inoue_Merz_Hikita_Raghu_Devereaux_Johnston_Hwang_2018} experiments reveal substantial spectral weight in ``replica'' bands, showing phonon effects beyond a simple quasiparticle mass renormalization.

\begin{figure}[h!]
    \centering
    \begin{subfigure}{\linewidth}
      \includegraphics[width=\linewidth]{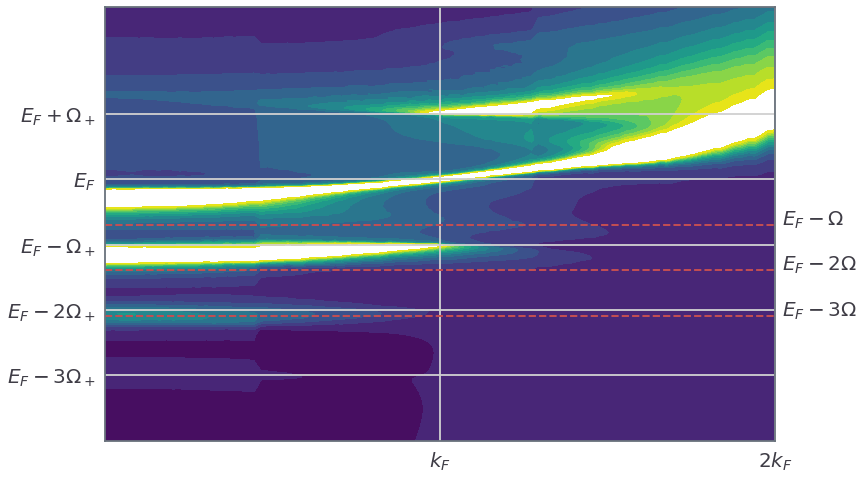}
      \caption{}
      \label{fig:519}
    \end{subfigure}\\
    \begin{subfigure}{\linewidth}
      \includegraphics[width=\linewidth]{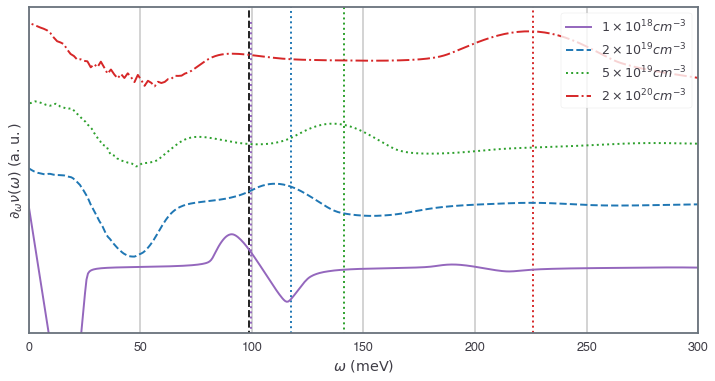}
      \caption{}
      \label{fig:dos}
    \end{subfigure}
    \caption{a) Spectral function $A(\mathbf{k},\omega)$ for \ce{SrTiO3} parameters with $n=5\times10^{19}\text{cm}^{-3}$. Solid and dashed horizontal lines are offset from the Fermi surface in multiples of the coupled mode frequency $\Omega_+$ and bare phonon frequency $\Omega$, respectively. b) First energy derivative of the computed density of states at different carrier densities. The vertical dashed line is the phonon energy, and the vertical dotted lines are $\Omega_+$ at each density.}
    \label{fig:mainfig}
\end{figure}

In this work we attempt to reproduce these observations by studying the normal-state properties of  the class of theories discussed above. We treat the Coulombic electron-phonon and electron-electron interactions on equal footing, and employ the cumulant expansion\cite{Kas_Rehr_Reining_2014} to incorporate the effects of multiple electron-boson interactions. Our main results are illustrated by the spectral function and density of states in Figure \ref{fig:mainfig}: although our formalism generates replica bands and qualitatively matches the experimental observation that they become weaker and more diffuse as the electron density is increased, we find that they occur at energies corresponding to the \emph{coupled} modes of the LO phonon and the collective response of the electron gas, which evolve with density. This is in glaring contrast to the experiments, which always find them fixed at the bare LO phonon energy.

\emph{Model} We treat the Hamiltonian $H = H_0 +H_\text{int}$ with
\begin{align}
\begin{split}\label{eq:hamiltonian}
    H_0 &= \sum_\mathbf{k}c^\dagger_\mathbf{k}\xi_\mathbf{k}c_\mathbf{k} +\Omega\sum_\mathbf{k}b^\dagger_\mathbf{k}b_\mathbf{k} \\
    H_\text{int} &= \sum_\mathbf{q}g_\mathbf{q}\rho_\mathbf{q}(b_\mathbf{q} +b^\dagger_\mathbf{-q}) +\sum_\mathbf{q}V_c(\mathbf{q})\rho_\mathbf{q}\rho_\mathbf{-q}.
\end{split}
\end{align}
$H_0$ describes a single band of electrons with isotropic dispersion $\xi_\mathbf{k} = k^2/2m -\mu$, where $\mu$ is the chemical potential, and an Einstein LO phonon with frequency $\Omega$. Although \ce{SrTiO3} has between one and three occupied conduction electron bands at experimentally accessible densities\cite{Marel_Mechelen_Mazin_2011,Lin_Fauque_Behnia_2015} and there is possibly some trace of their sequential filling in the superconducting phase diagram, inter-band scattering is large and superconductivity is single-gap\cite{Thiemann_Beutel_Dressel_Lee-Hone_Broun_Fillis-Tsirakis_Boschker_Mannhart_Scheffler_2018}, so we think a single band adequate to model the essential physics. There are likewise multiple phonons but the coupling of electrons to the LO mode near $\Omega = 100\mathrm{meV}$ is by far the strongest. This coupling is captured in the first, Fr{\"o}lich term of $H_\text{int}$, where $\rho_\mathbf{q} = \sum_\mathbf{k}c^\dagger_\mathbf{k}c_\mathbf{k+q}$ is the electron density operator at momentum $\mathbf{q}$ and the coupling $g^2_\mathbf{q} = \Omega\gamma\lambda/q^2$ is parameterized in terms of the long-range Coulomb coupling constant $\lambda = 4\pi e^2/\epsilon_\infty$ and a dimensionless parameter $\gamma$ which characterizes the stiffness of the TO modes that are present but do not couple to the electrons. The second term in $H_\text{int}$ is the repulsive electron-electron interaction, where $V_c = \lambda/q^2$ is the same long-range Coulomb coupling that appears in the Fr{\"o}lich term. In total this model is controlled by three parameters. The electron gas parameter $r_s = (\frac{4\pi}{3}n)^{-1/3}/a_B$ where $n$ is the density and $a_B = \frac{\hbar^2\epsilon_\infty}{me^2}$ is the effective Bohr radius, which we mean in the sense of the optical dielectric constant $\epsilon_\infty = 5.44$; the large static dielectric constant from the softening TO is an output in this formalism. \ce{SrTiO3} in this sense is quite dilute, with $r_s \sim 5$ at optimal doping, $n \sim 10^{20}\text{cm}^{-3}$ and as high as 40 at $n \sim 10^{17}\text{cm}^{-3}$, if such low-density samples are in fact uniform. The phonon frequency $\Omega_{\text{LO4}} = 100\text{meV}$ enters through an adiabaticity parameter $\Omega/E_F$, where $E_F$ is the Fermi energy; this parameter only drops into the adiabatic regime beyond $n=1.3\times 10^{17} \text{cm}^{-3}$. Finally, $\gamma$ ranges from $0$ for an uncoupled system to $\frac{1}{2}$ when the TO mode softens completely and can be thought of as a measure of proximity to ferroelectricity. In an ionic crystal with a rock salt structure, $\gamma = (1/\epsilon_\infty -1/\epsilon_0)/2$, but it is possible to straightforwardly generalize to the multi-phonon system, where $\sum_s \gamma_s$ plays the same role, and $\gamma_s$ can readily be extracted from e.g. reflectivity measurements for each LO mode $s$\cite{ToyozawaGreenBook}. We choose to use $\gamma=.4995$ from the measured values of $\epsilon_0$, as if the entirety of \ce{SrTiO3}'s large permittivity and electron-phonon coupling were attributable to a single LO-TO pair. A different choice does not qualitatively change our conclusions. We would like to emphasize that all of these parameters are in principle experimentally determined, and because the Fr{\"o}lich and Coulomb interactions are treated on an equal footing, there is no way within our model to independently tune the electron-phonon coupling strength without side-effects.

For comparison to experiments, we are interested in computing the spectral function at momentum $k$ and energy $\omega$
\begin{equation}
    A(\mathbf{k},\omega) = -\frac{1}{\pi}\text{Im} G^R(\mathbf{k},\omega)
\end{equation}
where $G^R$ is the retarded Green function. We begin by obtaining the leading-order effective electron-electron interaction,
\begin{equation}
    V_{\text{eff}}(\mathbf{q},\omega) = \frac{V_c(\mathbf{q}) +V_p(\mathbf{q},\omega)}{1 +\Pi(\mathbf{q},\omega)(V_c(\mathbf{q}) +V_p(\mathbf{q},\omega))}
    \label{eq:veff}
\end{equation}
where $V_p = 2g^2_\mathbf{q}\Omega/(\omega^2 -\Omega^2)$ is the phonon-mediated interaction and $\Pi(\mathbf{q},\omega)$ is the polarizability within the random phase approximation (RPA)\cite{giuliani_vignale_2005}. (\ref{eq:veff}) can equivalently be obtained by integrating out the phonons and electron density fluctuations to Gaussian order from (\ref{eq:hamiltonian}), performing a resummation of the bubble-type diagrams, or assuming assuming a dielectric function $\epsilon(\mathbf{q},\omega) = \epsilon_\infty +\epsilon_{\text{e-ph}} -V_c(\mathbf{q})\Pi(\mathbf{q},\omega)$ where $\epsilon_{\text{e-ph}} = (\epsilon_0 -\epsilon_\infty)/(1 -\omega^2/\Omega_{\text{TO}}^2)$ is the dieelctric function of a polar crystal with TO frequency determined by the Lyddane-Sachs-Teller relation $\Omega_{\text{LO}}^2/\Omega_{\text{TO}}^2 = \epsilon_0/\epsilon_\infty$\cite{mahan, Lyddane_Sachs_Teller_1941}.  

We next compute the one-loop ``$G_0W_0$'' self-energy,
\begin{equation}
    \Sigma_0(\mathbf{k},\omega) = \sum_{\mathbf{q}}\int\!\frac{d\omega^\prime}{2\pi}\, G_0(\omega +\omega^\prime, \mathbf{k} +\mathbf{q})V_{\text{eff}}(\mathbf{q},\omega^\prime)
\end{equation}
where $G_0(\mathbf{k},\omega) = (\omega -\xi_{\mathbf{k}} +i\delta)^{-1}$ is the free electron propagator. All computations are performed with $\delta = .001 E_F$. The integrals are split up just as in \cite{Hedin_1965}, and properties of this model for other semiconductors were already studied at this level by \cite{Kim_Das_Senturia_1978}. We next employ the retarded cumulant expansion of \cite{Kas_Rehr_Reining_2014} to obtain $G^R(\mathbf{k},\omega)$ by Fourier transforming after computing in the time domain
\begin{align}
\begin{split}\label{eq:cumulant}
    G^R(\mathbf{k},t) &= -i\theta(t)\exp(-i\tilde\xi_\mathbf{k} t)\exp(C^R(\mathbf{k},t)) \\
    C^R(\mathbf{k},t) &= \int\!\frac{d\omega}{\pi}\,\frac{|\text{Im}\Sigma_0(\mathbf{k}, \omega +\epsilon_\mathbf{k})|}{\omega^2}(e^{-i\omega t} -1)
\end{split}
\end{align}
where $\tilde\xi_\mathbf{k} = \epsilon_k +\text{Re}\Sigma_0(\mathbf{k},\epsilon_{\mathbf{k}})$ is the modified dispersion obtained from the one-loop calculation of $\Sigma_0$. Diagrammatically the cumulant expansion can be thought of as a partial resummation which uses the result of the one-loop self-energy to include rainbow diagrams and some crossing diagrams\cite{Gunnarsson_1994}. Although it is known to differ from more numerically accurate treatments of the electron gas for system sizes where those are available\cite{McClain_Lischner_Watson_Matthews_Ronca_Louie_Berkelbach_Chan_2015}, we chose this approximation for its one-shot generation of all replica bands (with spectral weight in each replica matching exact calculations in the limit of momentum-independent coupling to dispersionless bosons), and for its correct placement of those replicas relative to the renormalized bands, at minimal computational effort beyond $G_0W_0$. 

\emph{Results and Discussion} In Figure \ref{fig:519} we plot the spectral function for \ce{SrTiO3} at $n=5\times10^{19}\text{cm}^{-3}$. All spectral functions are plotted on a logarithmic color scale to enhance the visibility of faint features. Near the Fermi surface, the quasiparticle band is narrow with typical $\text{Im}\Sigma \sim \omega^2$ Fermi-liquid behavior. At a density-dependent momentum away from the Fermi surface, there is a sudden increase in broadening which produces a characteristic kink. Below the quasiparticle band there is a series of replica bands of similar width near the bottom of the band, and vanishing as $k\to k_F$ and more spectral weight is transferred to the quasiparticle. The replicas are spaced by an energy $\Omega_+ \neq \Omega$, which can be understood by considering the large-$r_s$ plasmon pole limit, when $\Pi(\mathbf{q},\omega) \to \Omega_p^2(\mathbf{q})/(\omega^2 V_c(\mathbf{q}))$ where $\Omega_p^2$ is the plasmon frequency. $V_\text{eff}$ then has poles at
\begin{equation}
    \Omega^2_\pm = \frac{1}{2}\left(\Omega^2 +\Omega_p^2 \pm \sqrt{\Omega^4 +\Omega_p^4 +(8\gamma -2)\Omega^2\Omega_p^2}\right).
    \label{eq:coupledmodes}
\end{equation}

\begin{figure}
    \centering
    \begin{subfigure}[b]{.5\linewidth}
      \includegraphics[width=\linewidth]{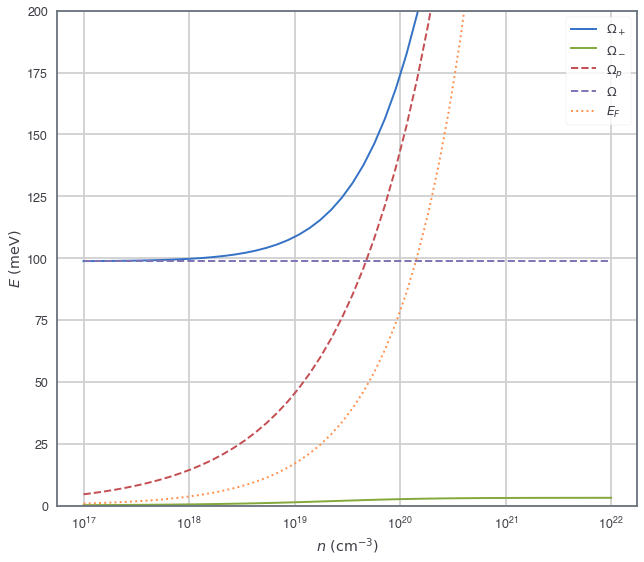}
      \caption{}
      \label{fig:guideplot_mev}
    \end{subfigure}\hfill
    \begin{subfigure}[b]{.5\linewidth}
      \includegraphics[width=\linewidth]{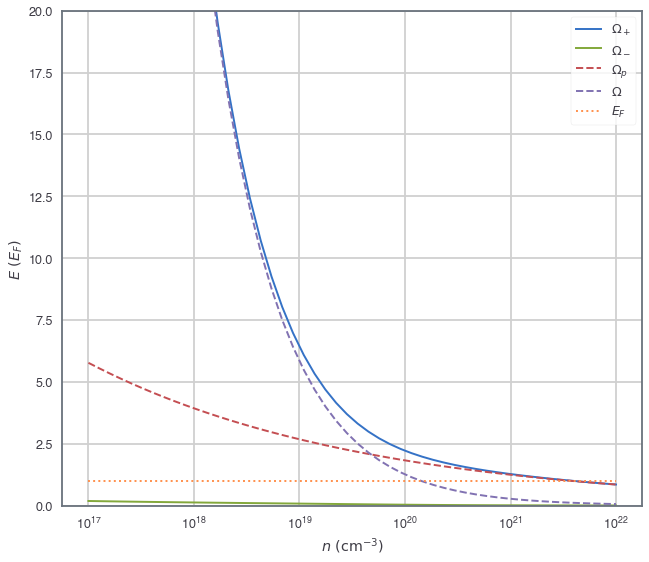}
      \caption{}
      \label{fig:guidefig_ef}
    \end{subfigure}
    \caption{Coupled modes of (\ref{eq:coupledmodes}) (solid), along with uncoupled modes (dashed) and $E_F$ (dotted), in physical units (left) and in units of $E_F$ (right)}
    \label{fig:coupledmodes}
\end{figure}

We plot these modes at $\mathbf{q}=0$ alongside the bare phonon and plasmon for \ce{SrTiO3} in Figure \ref{fig:coupledmodes}. For small and intermediate $\gamma$ the coupled modes largely follow the bare modes except in the vicinity of the anti-crossing, but as $\gamma$ approaches its maximum value of $1/2$, as in the case of \ce{SrTiO3}, $\Omega^2_+ \to \Omega^2 +\Omega_p^2$ while $\Omega_- \to 0$. In this limit the oscillator strength in the lower mode vanishes as $\Omega_-^3$. The lower coupled mode is therefore too weak for us to observe with \ce{SrTiO3} parameters and we only see signatures of $\Omega_+$. This represents a major discrepancy between the model and experimental observations such as the photoemission experiment of \cite{Wang_Walker_Tamai_Wang_Ristic_Bruno_Torre_Ricco_Plumb_Shi_2016}, which finds replica bands at the phonon frequency at all dopings. 

For comparison to the tunneling experiments of \cite{Swartz_Inoue_Merz_Hikita_Raghu_Devereaux_Johnston_Hwang_2018}, from the spectral function we may also compute the density of states, $\nu(\omega) = \sum_{\mathbf{k}} A(\mathbf{k},\omega)$. At high densities, $\nu$ broadly follows the $\sqrt{\omega}$ free-particle prediction, with small deviations due to the spectral weight in the replica bands, which become increasingly prominent and eventually dominant as the density is decreased. To focus on these features, in Figure \ref{fig:dos} we plot $d\nu/d\omega$ for densities from $10^{18}-2\times10^{20}\text{cm}^{-3}$. With increasing density these evolve from a sharp peak-dip structure to a broad peak, matching what is seen in experiment, but at all densities they remain centered at $\omega = \Omega_+$, again at odds with the experimental observation of replicas at the phonon frequencies, independent of density.

\begin{figure}
    \includegraphics[width=\linewidth]{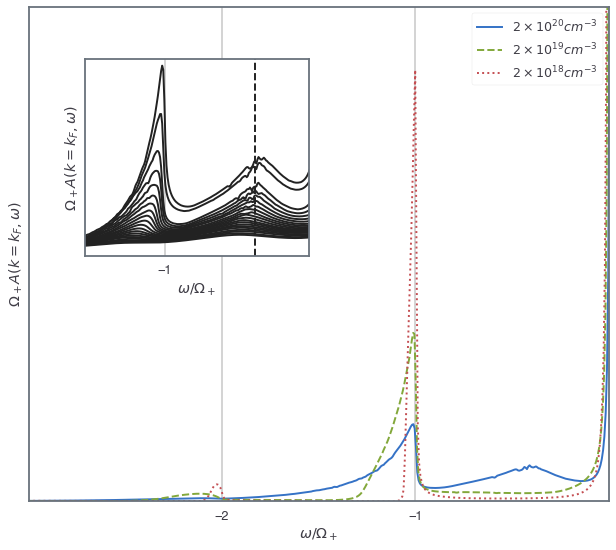}
    \caption{Fermi surface spectral function $A(\mathbf{q} = k_F, \omega)$ below $E_F$ at different densities showing the evolution of the replica bands. Inset: $n=2\times10^{20}\text{cm}^{-3}$ spectral functions from $k=k_F$ (top) to $k=1.4k_F$ (bottom), showing the evolution of dispersionless features at $\Omega_+$ and $\Omega$ (dahsed line).}
    \label{fig:comboplot}
\end{figure}

A striking feature in the photoemission data is the disappearance of replica bands at high densities, and the development of a quasiparticle band with a kink. Similar phenomenology in \ce{TiO2} was recently explained in terms of Thomas-Fermi screening of the electron-phonon matrix element\cite{Verdi_Caruso_Giustino_2017}. In Figure \ref{fig:comboplot} we show a cut of $A(k=k_F,\omega)$ with frequencies scaled by $\Omega_+$ for three different densities. The effects of screening are most apparent in the decreased amplitude and greater broadening of the replica peaks with increasing density. The screening wavevector $q_{TF}$ grows with density and enters the electron-boson matrix element as $1/(q^2 +q_{TF}^2)$, increasing the range of momenta around $q=0$ which contribute significantly and thereby broadening the peak. This effect is insufficient to qualitatively modify the spectrum, however.

A qualitative difference that does emerge at high densities is a broad feature between the quasiparticle and the first replica band. We study this feature as a function of momentum $k>k_F$ in the inset of Figure \ref{fig:comboplot}, and find that it is nearly non-dispersing and centered around the bare phonon frequency, $\Omega$. (A second feature is seen at $\omega=\Omega_+$, and at all the densities we have studied we observe non-dispersing tails of all replica bands at $k>k_F$. \cite{Kas_Rehr_Reining_2014} has found a similar effect in the homogeneous electron gas.)  At this density, $\Omega/E_F < 1$; this feature disappears as one moves into the anti-adiabatic regime, although it is still faintly present in Figure \ref{fig:519}, where $\Omega/E_F \sim 2$. The origin of this feature may be understood by examining the structure of the dielectric function $\epsilon(\mathbf{q},\omega)$. In the $q$-$\omega$ plane, it inherits from the electron gas the well-known electron-hole continuum, bounded by $\omega_{\pm}(q) = q^2/2m +v_F q$ where $\text{Im}\epsilon \neq 0$ \cite{giuliani_vignale_2005}. To the left of $\omega_-$, $\text{Im}\epsilon$ is nonzero along the dispersions of the coupled modes discussed above, but after the plasmon disperses into the continuum, only a single peak at the bare phonon frequency emerges beyond $\omega_+$. We interpret the feature, positioned in a region where $\Im\epsilon =0$ at the one-loop level, as a multiple-scattering process involving this mode and the electron-hole continuum.

\begin{figure}
    \centering
    \begin{subfigure}[b]{.5\linewidth}
      \includegraphics[width=\linewidth]{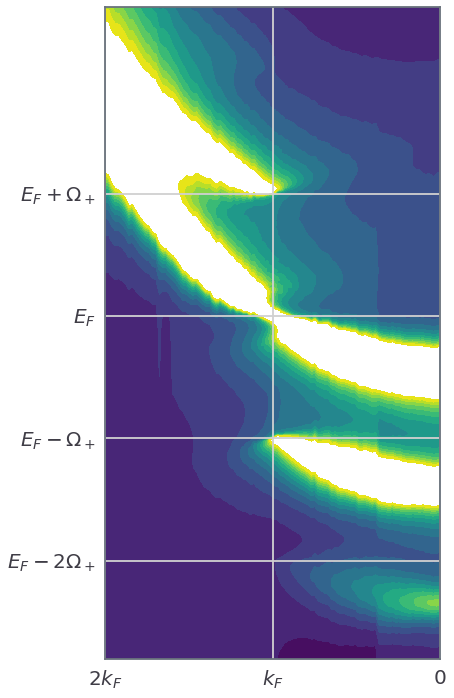}
      \caption{}
      \label{fig:o05}
    \end{subfigure}\hfill
    \begin{subfigure}[b]{.5\linewidth}
      \includegraphics[width=\linewidth]{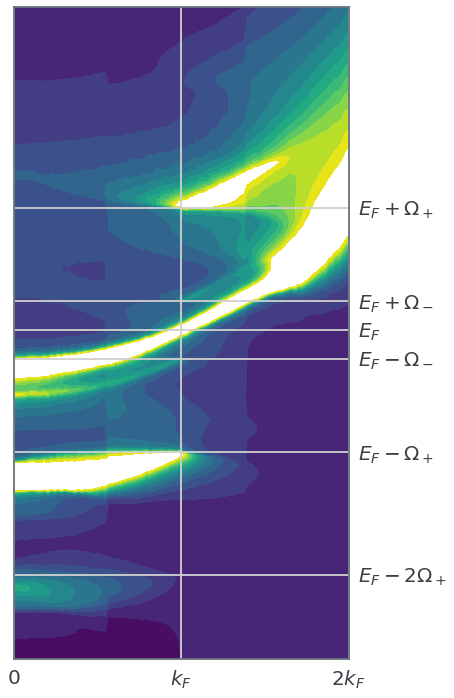}
      \caption{}
      \label{fig:g4}
    \end{subfigure}
    \caption{Like Figure \ref{fig:519}, but with altered parameters a) $\Omega/E_F = .5$, b) $\gamma = .4$. }
    \label{fig:params}
\end{figure}

To further explore the importance of adiabaticity within our model, in Figure \ref{fig:params} we artificially vary $\Omega/E_F$ and $\gamma$ away from \ce{SrTiO3} parameters while holding the density fixed at $n=5\times 10^{19}\text{cm}^{-3}$. In Figure \ref{fig:o05} the phonon is made adiabatic, $\Omega/E_F = .5$. The quasiparticle and replica bands are substantially broadened, the kink in the quasiparticle moves much closer to the Fermi surface, and the nondispersing feature $\Omega$ below $E_F$ becomes much stronger compared to Figure \ref{fig:519}. In Figure \ref{fig:g4}, we instead set $\gamma=.4$, substantially increasing the frequency and oscillator strength of $\Omega_-$; note, however, that still $\Omega_- < E_F$ at this density. In this case, an additional, narrow replica band emerges offset from the quasiparticle band, which maintains a modest linewidth with no kinks to the bottom of the band. (A very faint non-dispersing feature at the bare phonon frequency is still detectable, if difficult to resolve by eye.) Here it is worth noting the theory of \cite{Ruhman_Lee_2016}, which suggests that a modest reduction in $\epsilon_0$ is sufficient to impart sufficient strength to the $\Omega_-$ mode to mediate pairing while remaining adiabatic. To reproduce the phenomenology of Figure \ref{fig:g4}, however, with $\Omega_- \approx 25\mathrm{meV}$, requires a far greater reduction of $\epsilon_0 \approx 30$, and we note that in this case $\Omega_+$ remains strongly shifted away from $\Omega$.

These results collectively suggest that the transition between replica bands in the spectral function and a broadened quasiparticle with kinks is driven by the interplay of adiabaticity and screening by the electron-hole continuum. The latter imparts some intrinsic width to the quasiparticle band and causes the electron-boson matrix element to deviate from the forward-scattering limit, broadening any replica bands. If the width of the quasiparticle band exceeds the  energy difference between it and the replica, the entire structure is absorbed into a broadened quasiparticle. At some point as the intrinsic quasiparticle width decreases as it approaches the Fermi surface, this condition will be violated, causing a kink as the quasiparticle band suddenly narrows. In the case of coupled modes we have studied here, spectral weight at the Fermi surface is not exclusively distributed between the quasiparticle and any remaining replicas, but may also accumulate in a broad feature around the bare phonon frequency. 

\emph{Conclusion} Our model neglects many experimentally significant features of \ce{SrTiO3}, including the occupation of multiple electron bands and the coupling to mulitple optic phonons. While the inclusion of accurate electron and phonon band structures is necessary to obtain quantitative agreement with experiment, we do not believe it would be sufficient to resolve the major discrepancy we observe here, namely, that the normal-state spectral features within our model are predicted to be unambiguously shifted away from the bare phonon frequencies through hybridization with the plasmon. This presents a challenge to a class of theories of superconductivity which rely on the existence of these hybrid modes, or more broadly plasmon-mediated superconductivity theories dating back to \cite{Grabowski_Sham_1984}.

Equivalently, our results suggest that future work investigate the nature of the plasmon in \ce{SrTiO3}, which has been indirectly observed in reflectivity data to have a width of the same order as the plasma frequency \cite{Gervais_Servoin_Baratoff_Bednorz_Binnig_1993}. Heuristically, this is consistent with other observations that suggest the electron gas in \ce{SrTiO3} behaves as if it were directly screened by $\epsilon_0$\cite{Rischau_Lin_Grams_Finck_Harms_Engelmayer_Lorenz_Gallais_Fauque_Hemberger_2017}, or equivalently that its diluteness in terms of $r_s$ is unremarkable compared to other semiconductors. If such screening exists, it is a more elusive effect than the multiple-scattering processes we have considered here.

Our theory ultimately suffers from a defect similar to that found in many of the superconducting theories: it is formally an expansion in $r_s$, which is always substantially greater than unity at the densities we have considered. Paradoxically, as the theory becomes more poorly controlled at low densities, the phenomenology becomes simpler as the response is totally dominated by the plasmon pole physics of (\ref{eq:coupledmodes}). More sophisticated techniques are needed to understand the interplay of polaron physics and the many-electron fluid.

\bibliography{../e-ph.bib}

\end{document}